\documentclass[10pt]{article}
\usepackage[left=4cm, right=3cm, top=2cm, bottom=2cm]{geometry}
\usepackage{authblk}
\usepackage[tbtags]{amsmath}
\usepackage{bm}% bold math
\usepackage[onehalfspacing]{setspace}

\usepackage{textcomp,gensymb}
\usepackage{quoting}
\usepackage{epsfig}
\usepackage{color}
\usepackage{amsmath}
\usepackage{cite}
\usepackage{graphicx}
\usepackage{subfig}
\usepackage{anyfontsize}
\usepackage{t1enc}

\begin{document}

\title{\textbf{Generalized Backpropagation Algorithms for Diffraction Tomography}}

\author{Pavel~Roy~Paladhi%
\thanks{E-mail: \texttt{roypalad@egr.msu.edu}; Corresponding author}}
\affil{\textit{{\fontsize{10}{10}\selectfont Department of Electrical and Computer Engineering, Michigan State University, 428 S Shaw Lane, Rm-2120, East Lansing, Michigan}}}

\vspace{1in}

\author{Ashoke~Sinha$^{\dagger}$, Amin~Tayebi$^{\S,\ddagger}$, and Lalita~Udpa$^{\S}$}
\affil{\textit{{\fontsize{10}{10}\selectfont$^\dagger$Department of Statistics and Probability, Michigan State University, 619 Red Cedar Road, C413 Wells Hall, East Lansing, Michigan 48824, USA.}}}
\affil{\textit{{\fontsize{10}{10}\selectfont $^{\S}$Department of Electrical and Computer Engineering, Michigan State University, 428 S Shaw Lane, Rm-2120, East Lansing, Michigan}}}
\affil{\textit{{\fontsize{10}{10}\selectfont $^{\ddagger}$Department of Physics and Astronomy, Michigan State University, 567 Wilson Rd, East Lansing, Michigan}}}

\date{Dated: \today}

\maketitle

(A shorter version of this paper is published in Progress In Electromagnetics Research B. Vol. 66)

\begin{abstract}
  Filtered backpropagation (FBPP) is a well-known technique used for Diffraction Tomography (DT). For accurate reconstruction of a complex image using FBPP, full $360^{\circ}$ angular coverage is necessary. However, it has been shown that using some inherent redundancies in projection data in a tomographic setup, accurate reconstruction is still possible with $270^{\circ}$ coverage which is called the minimal-scan angle range. This can be done by applying weighing functions (or filters) on projection data of the object to eliminate the redundancies and accurately reconstruct the image from this lower angular coverage. This paper demonstrates procedures to generate many general classes of these weighing filters. These are all equivalent at $270^{\circ}$ coverage but would perform differently at lower angular coverages and under presence of noise. This paper does a comparative analysis of different filters when angular coverage is lower than minimal-scan angle of $270^{\circ}$. Simulation studies have been done to find optimum weight filters for sub-minimal angular coverage. The optimum weights can generate images comparable to a full $360^{\circ}$ coverage FBPP reconstruction.  Performance of these in presence of noise is also analyzed. These algorithms are capable of reconstructing almost distortionless complex images even at angular coverages of $200^{\circ}$.
\end{abstract}

\section{Introduction}

Diffraction tomography (DT) is a popular imaging technique \cite{devaney1982filtered}-\cite{tabbara1988diffraction} applied to a variety of applications like medical imaging, non-destructive evaluation of materials, structural health monitoring, geophysics etc. \cite{schueler1984fundamentals}-\cite{ritzwoller_global_surfwav_DT}. In the domain of optical imaging this method  has been  explored in depth viz. optical diffraction tomography (ODT) with multi-disciplinary applications \cite{gorski_fbpp_phot_crys}-\cite{kostencka_odt_capil_odt}. DT is a broad imaging technique of which  ultrasound and microwave tomographic imaging are sub-classes. It is a comprehensive way of characterizing the complex valued object-function of the test object. The scheme is applicable to microwave tomography of tissue samples. It has been of great interest in tomography of human breast to identify malignant tumours \cite{catapano_quant_mwt}-\cite{paladhi_qnde2014}. The high contrast between healthy and cancerous breast tissue provides opportunities for clearly identifying malignancies within breast tissue. The contrast is much higher than in case of X-rays and hence microwave tomography is advantageous both in terms of true detection and radiation damage from X-rays.  Further applications are in fields of SAR imaging. Various backprojection techniques have been explored for use in radar imaging and modified for better reconstructions \cite{ren_3d_sar_tom}-\cite{capozzoli_gpu_sar_backproj}. Faster implementations for backprojections algorithms are also being intensely explored e.g. \cite{capozzoli_gpu_sar_backproj}. Again, radar based methods have been combined with microwave tomography to generate higher resolutions for medical imaging \cite{baran_brst_mwt_rdr_based_est}. Thus, any improvement in the traditional backpropagation techniques is of potential interest across multiple disciplines. In the presence of weak scatterers, assuming the Born or Rytov approximations \cite{avinash2001principles}, \cite{kaveh1979ultrasonic}, \cite{kaveh1984signal}, the Fourier Diffraction Projection theorem (FDP) can be applied. The FDP relates the scattered field data from the Region of Interest (ROI) due to incident plane waves to the 2D-Fourier Space of the ROI, i.e. the test or sample object. Using this image reconstruction is possible from collecting scattered field data from the ROI.

Devaney developed the famous filtered backpropagation (FBPP) algorithm for this configuration which reconstructed a low-pass filtered image of the object function \cite{devaney1982filtered}, \cite{devaney1983computer}. The traditional backpropagation technique requires projection data from $[0, 2\pi]$ angular coverage for accurate image reconstruction of a complex image. In \cite{pan1999minimal}, \cite{pan1998unified}, \cite{anastasio2001full}, \cite{pan2002limited} it was shown that using inherent redundancies in the projection data from the conventional setup, exact reconstructions were possible with data from $[0,  3\pi/2]$ coverage. In this paper we explore a technique such that the redundancies in the Fourier space data from the conventional setup can be effectively used over any range between $\pi$ to $ 3\pi/2$ to generate better reconstruction over regular backpropagation. This is a direct reconstruction method which does not employ any error minimization algorithms and hence is faster and accurate over angular coverages between $180^{\circ}$-$270^{\circ}$. This is very crucial because the major challenge in many real world situations is that projection measurements cannot be gathered over full $360^{\circ}$ view around the test object. With limited access to the object and decrease in angular coverage, the available data in the Fourier space decreases. Reconstruction from this partially known Fourier space leads to many artifacts and loss of important image features. If standard FBPP algorithms are used for limited angular coverage, the reconstruction quickly deteriorates. Hence alternate schemes are needed for image reconstruction from limited angular coverage projection data.  This paper proposes to use projection datasets optimally to get better reconstructions from lower angular coverages. The idea was briefly proposed in \cite{roy_qnde2015} and \cite{roy_aces2015}. A shorter version of this article is published in \cite{roy2016improved}.

The paper is arranged as follows: Section \ref{fdp} gives a short background of the FDP and FBPP algorithms. Section \ref{mincomp} explains the minimal-scan complete dataset proposed in \cite{pan1999minimal} and how equivalent systems of backpropagation algorithms can be generated. Section \ref{wtgen} introduces methods to generate optimized classes of backpropagation algorithms which can give better reconstruction than regular FBPP in the angular coverage range where the redundancy can still be exploited, i.e. between $180^{\circ}$-$270^{\circ}$. Section \ref{result} presents results from different backpropagation classes and analyses the relative performances at angular coverages below $270^{\circ}$.

\section{FDP \& FBPP} \label{fdp}

\begin{figure}[!ht]
  \centering
  \includegraphics[width=0.75\textwidth]{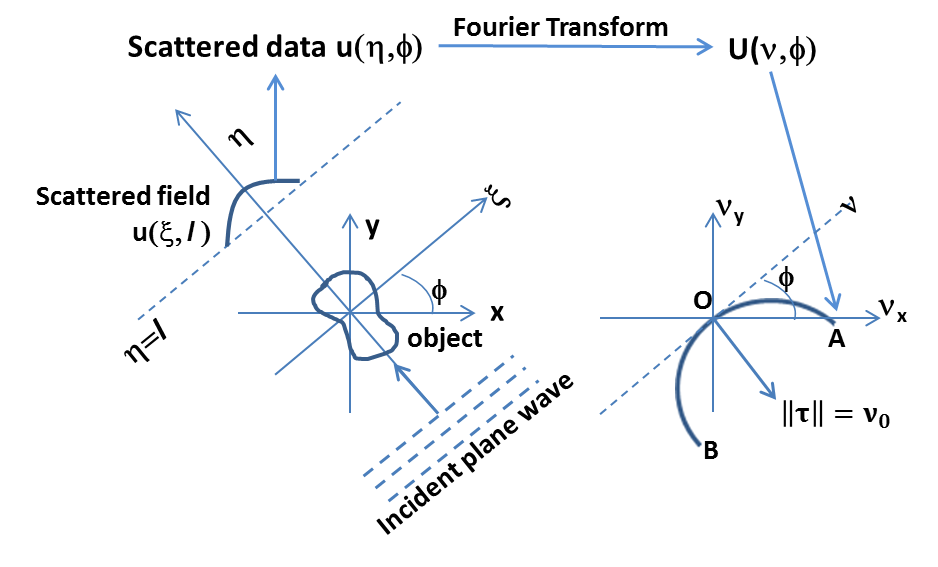}
  \caption{\label{2dft} Classical scan configuration of 2D DT (left) and relation of scattered field data with the 2D Fourier space of the objective function (right).}
\end{figure}

The fundamental theory behind 2D-DT is the FDP which relates the scattered field data from the Region of Interest (ROI) due to incident plane waves to the 2D-Fourier Space of the ROI, i.e. the test or sample object \cite{avinash2001principles}, \cite{mueller1979reconstructive}. The traditional 2D-configuration is shown in Fig. \ref{2dft}. If the object $o(x,y)$ is illuminated with a monochromatic plane wave of frequency $\nu_0$ incident at an angle $\phi$ to the horizontal axis, the 1D Fourier Transform (FT) of the scattered field measured along the straight line $\eta=l$ in the co-ordinate system $(\xi,\eta)$ gives the values of the 2D transform of the object $O(\nu_x,\nu_y)$ along a semi-circular arc in the frequency domain tilted at angle $\phi$ as shown in the right half of Fig. \ref{2dft}. The scattered field data and the object function are related by the following equation:
\begin{equation}
  U_B(\nu,l) = \frac{j}{2\sqrt{\nu_0^2-\nu^2}} e^{j\sqrt{\nu_0^2-\nu^2 l}}O\left(\nu, \sqrt{\nu_0^2-\nu^2}-\nu_0\right),
\end{equation}
where $U_B(\nu,l)$ represents 1D FT of the scattered field under Born approximation (measured at line $\eta=l$), and $\nu$ ranges between $[-\nu_0, \nu_0]$.

Devaney developed the well-known filtered backpropagation method \cite{devaney1982filtered}. In polar coordinates, as presented in \cite{pan1999minimal}, the backpropagation integral takes the form:
\begin{equation}
\begin{split}
  a(r, \theta) = \frac{1}{2} \int_{\phi=0}^{2\pi} & \int_{\nu_m=-\nu_0}^{\nu_0} \frac{\nu_0}{\nu'} |\nu_m| M(\nu_m,\phi) \\
  & \cdot e^{[j2\pi \nu_m \cos(\phi-\alpha-\theta)]} d\nu_md\phi, \label{aeqn}
\end{split}
\end{equation}
where, $a(r,\theta)$ is the objective function being reconstructed in polar spatial coordinates $(r,\theta)$, $\nu_0$ is the frequency of the incident monochromatic plane wave of frequency $\nu_0$ and $\phi$ being the incidence angle, $M(\nu_m,\phi)$ is  a modified 1D FT of the scattered data, defined as
\begin{equation}
  M(\nu_m, \phi) = U_B(\nu_m, \phi) \frac{j\nu'}{2\pi^2\nu_0^2} e^{-j2\pi\nu'l}, \tag{2a}
\end{equation}
where \qquad $\nu' = \sqrt{\nu_0^2-\nu_m^2}$, \qquad $\nu_a = \text{sgn}(\nu_m) \sqrt{\nu_m^2-\nu_{\mu}^2}$, \qquad $\nu_{\mu} = j(\nu'-\nu_0)$ and
\begin{equation}
  \alpha = \text{sgn}(\nu_m) \arcsin\left( \frac{1}{2\nu_0}\sqrt{\nu_m^2-\nu_{\mu}^2} \right), \tag{2b}
\end{equation}
Without loss of generality, $\alpha$ could be further simplified to
\begin{equation}
  \alpha = \frac{1}{2} \arcsin\left(\frac{\nu_m}{\nu_0}\right). \tag{2c}\label{alphaeq}
\end{equation}
To reconstruct a complex object accurately, a full knowledge of $M(\nu_m,\phi)$ in the range $[0, 2\pi]$ is necessary to perform the integration of (\ref{aeqn}) - see \cite{devaney1982filtered}, \cite{devaney1989limited}. The next section explores how this requirement can be brought down to an angular coverage of $[0,  3\pi/2]$.

\section{The Minimal Complete Coverage in DT} \label{mincomp}

Consider the two semi-arcs OA and OB from fig-1. The two arcs individually traverse the transform space (i.e. the Fourier space) as the interrogating wave angle changes between $[0, 2\pi]$. Now consider their traversals in the Fourier space individually for angular coverage of $[0,  3\pi/2]$. This is illustrated in Fig. \ref{rotate} below. Individually each half does an incomplete traversal of the Fourier space as shown in Fig. \ref{rotate}(a) and \ref{rotate}(b). However, if superimposed, as seen in Fig. \ref{rotate}(c), they in fact, traverse the entire Fourier space with some areas of overlap. The key point is that at $270^{\circ}$ there is in effect, a complete coverage of the Fourier space. This scan range of $[0, 270^{\circ}]$ is referred to as the minimal scan angle and is the angular coverage required for exact reconstruction (see \cite{pan1999minimal}).

\begin{figure}[!ht]
  \centering
   \includegraphics[width=0.75\textwidth]{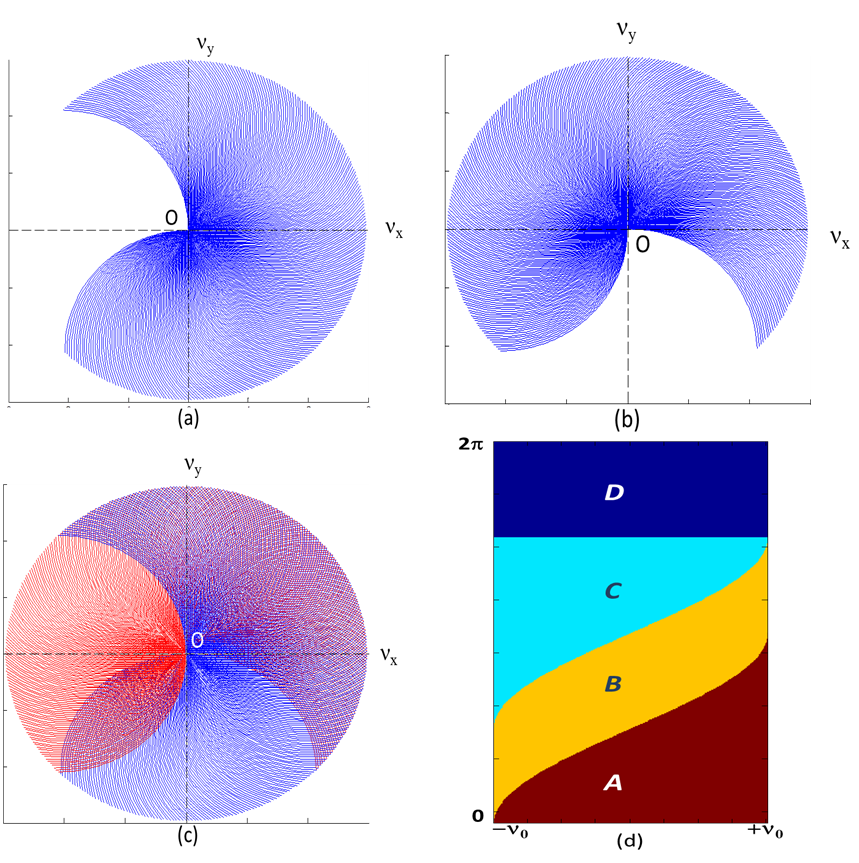}
  \caption{\label{rotate}(a) Fourier space coverage for $270^{\circ}$ angular access by segment OB in fig.1 (b) same for segment OA of fig-1. (c) Superposition of the two coverages from (a) and (b). (d) The fourier data-space in alternate co-ordinate system with angular coverage along y-axis and the wave number along x-axis.}
\end{figure}

To better appreciate the principle, the Fourier space domain is re-plotted in a modified coordinate system, where the spatial frequency is plotted along $x$-axis and the angular coverage along $y$-axis. Physically this new coordinate system can be viewed as `straightening' the arcs $AOB$ from each projection and stacking them up on top of each other sequentially.  In this layout, the entire Fourier dataspace can be divided into four sub-regions $A,B,C$ and $D$ as shown in Fig. \ref{rotate}(d). The boundaries for the four regions can be expressed as  $A=[|\nu_m | \leq \nu_0, 0 \leq \phi < 2\alpha +  \pi/2]$, $B=[|\nu_m | \leq \nu_0,  \pi/2 + 2\alpha \leq \phi < 2\alpha + \pi]$,  $C=[|\nu_m | \leq \nu_0, \pi+2\alpha \leq \phi <  3\pi/2]$ and $D=[|\nu_m | \leq \nu_0, 3\pi/2 \leq \phi < 2\pi]$. As seen in (\ref{alphaeq}), $\alpha$ is a function of $\nu_m$ and so the regions have nonlinear boundaries.

\begin{figure}[!ht]
\centering
  \includegraphics[width=0.75\textwidth]{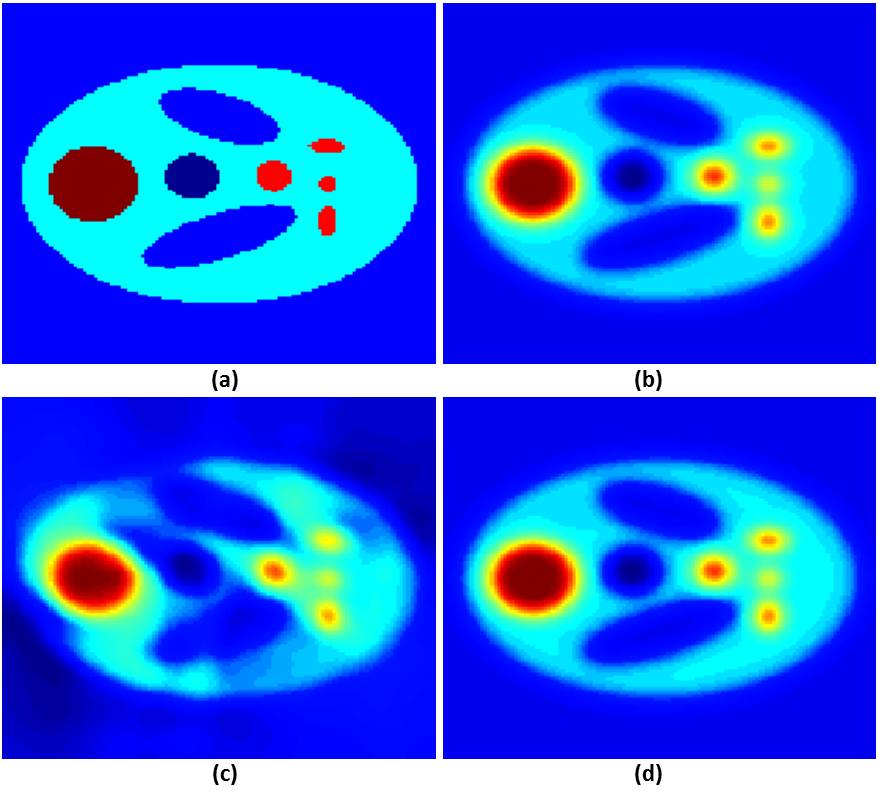}
  \caption{\label{fbppReal} Demonstration of the MS-FBPP concept (a) real part of original image, (b) FBPP reconstruction from $360^{\circ}$ coverage (c) FBPP reconstruction from $270^{\circ}$ coverage and (d) MS-FBPP reconstruction from $270^{\circ}$ coverage}
\end{figure}

\begin{figure}[!ht]
\centering
  \includegraphics[width=0.75\textwidth]{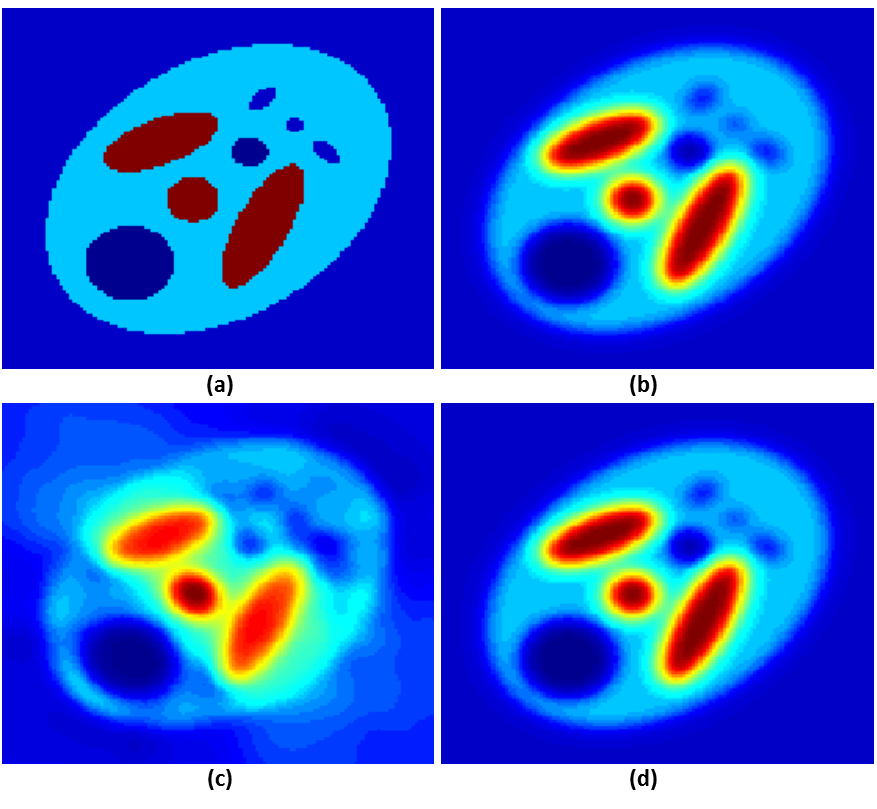}
  \caption{\label{fbppImg} Demonstration of the MS-FBPP concept (a) imaginary part of original image, (b) FBPP reconstruction from $360^{\circ}$ coverage (c) FBPP reconstruction from $270^{\circ}$ coverage and (d) MS-FBPP reconstruction from $270^{\circ}$ coverage}
\end{figure}

From FDP theorem the following periodicity can be shown (see \cite{pan1998unified}):
\begin{equation}
  M(\nu_m,\phi)=M(-\nu_m,\phi+\pi-2\alpha),
\end{equation}
which also implies that for every point in region $A$ (respectively $B$), there is an point of identical value in region $C$ (respectively $D$). Thus, the knowledge of $M(\nu_m,\phi)$ in regions $A$ and $B$, makes information in $C$ and $D$ redundant. This redundancy can be handled by normalizing the dataspace appropriately using appropriate weight filters. The weighted dataset $M'(\nu_m,\phi)$ can be defined as $M'(\nu_m,\phi)= w(\nu_m,\phi)M(\nu_m,\phi)$, where $w(\nu_m,\phi)$ satisfies the condition
\begin{equation}
   w(\nu_m,\phi)+w(-\nu_m,\phi+\pi-2\alpha) = 1. \label{wt1}
\end{equation}
It should be noted that since region D is unavailable in a $270^{\circ}$ coverage we set $w(\nu_m,\phi)=0$ in region D and correspondingly $w(\nu_m,\phi)=1$ in region B. For regions A and C, any weight functions are valid as long as they satisfy (\ref{wt1}). Using the weighted dataset $M'(\nu_m,\phi)$ we can apply the regular backpropagation algorithm for reconstruction. This is called Minimal-Scan Filtered Backpropagation (MS-FBPP) which involves evaluating the following integral
\begin{equation}
\begin{split}
     a^W(r, \theta) = \frac{1}{2} \int_{\phi=0}^{ 3\pi/2} & \int_{\nu_m=-\nu_0}^{\nu_0} \frac{\nu_0}{\nu'} |\nu_m| M(\nu_m,\phi) \\
     & \cdot e^{[j2\pi \nu_m \cos(\phi-\alpha-\theta)]} d\nu_md\phi.
\end{split}
\end{equation}
This integral can in theory exactly reconstruct an image from a $270^{\circ}$ angular coverage by utilizing data redundancy in projection data. The weights $w(\nu_m,\phi)$ can then be used to define classes of backpropagation algorithms for image reconstruction. An example of weight functions introduced in \cite{pan1999minimal} is given in equation below:
\begin{equation}
  w(\nu_m, \phi) =
  \begin{cases}
    \sin^2\left[ \frac{\pi}{4} \cdot \frac{\phi}{\pi/4+\alpha}\right], & \quad \text{in } A, \\
    1, & \quad \text{in } B, \\
    \sin^2\left[ \frac{\pi}{4} \cdot \frac{ 3\pi/2 - \phi}{\pi/4-\alpha}\right], & \quad \text{in } C, \\
    0, & \quad \text{in } D.
  \end{cases} \label{Panwt}
\end{equation}
This will be used as a reference later on in Section \ref{result} for comparison with the weight functions that are proposed in this paper. The filter of (\ref{Panwt}) will be referred to as sine-squared filter later in this paper. This is a very elegant and simple filter designed to have continuity across the boundaries between regions $A, B, C$ and $D$. To demonstrate the efficacy of MS-FBPP, a sample reconstruction is performed on a complex image. Reconstruction from regular FBPP and MS-FBPP using weights of (\ref{Panwt}) are given in Fig. \ref{fbppReal} and Fig. \ref{fbppImg}. In practice, the real and imaginary components maybe expected to look very similar as the parameter distributions would be determined by tissue and organ boundaries, however in this paper, the real and imaginary components of the test phantom have been kept different to retain a generalized approach. The images show the MS-FBPP algorithms capable of generating accurate reconstructions from $270^{\circ}$, equivalent to full coverage FBPP whereas the regular FBPP image shows considerable distortion at $270^{\circ}$.

Weight filters which are discontinuous across boundaries can give rise to artifacts especially in case of discrete data. Further, for each class of weights, its distribution in frequency space of Fig. \ref{rotate}(d) also determines the performance when the available coverage is below $270^{\circ}$, because, $C$ and $D$ are complementary to regions $A$ and $B$ respectively. An efficient weight function set would be that which spans most of the regions $A$ and $B$, thus limiting the requirement to access regions $C$ and $D$. In effect such weights can generate good reconstructions even from angular coverages below $270^{\circ}$. The next section explains a systematic approach to generate general classes of weight functions.

\section{Generation of Weight Function Classes} \label{wtgen}

From previous section, we see that the weights in regions $B$ and $D$ are $1$ and $0$ respectively. Notice that weights in region $C$ can be generated from weights of region $A$, because for any point $(\nu_m, \phi)$ in region $C$, the point $(-\nu_m,\phi+\pi-2\alpha)$ is in $A$, and then using (\ref{wt1}), we get $w(\nu_m,\phi)= 1 - w(-\nu_m,\phi+\pi-2\alpha)$. Thus it is sufficient to generate weights for the region $A$ only.

Furthermore from (\ref{wt1}), the non-negative weights are bounded above by 1. So in region $A$, for any fixed $\nu_m$, the function $w(\nu_m, \cdot) =: F(\cdot)$ is defined on $[0, 2\alpha+\pi/2]$, and takes values between $0$ and $1$. Hence we propose a generalized approach by using cumulative distribution functions (cdf), which are guaranteed to be bounded within $0$ and $1$. Though any cdf will serve this purpose, we would prefer to have cdf's with support in $[0, 2\alpha+\pi/2]$ for any fixed $\nu_m \in [-\nu_0, \nu_0]$, because that will ensure that weights are continuous at the boundary between regions $A$ and $B$, and consequently at the boundaries between regions $B$ and $C$, and $C$ and $D$. Weights which are discrete at the boundaries will generate artifacts in case of discrete datasets \cite{pan1999minimal}.

\begin{figure}[!ht]
\centering
  \includegraphics[width=0.75\textwidth]{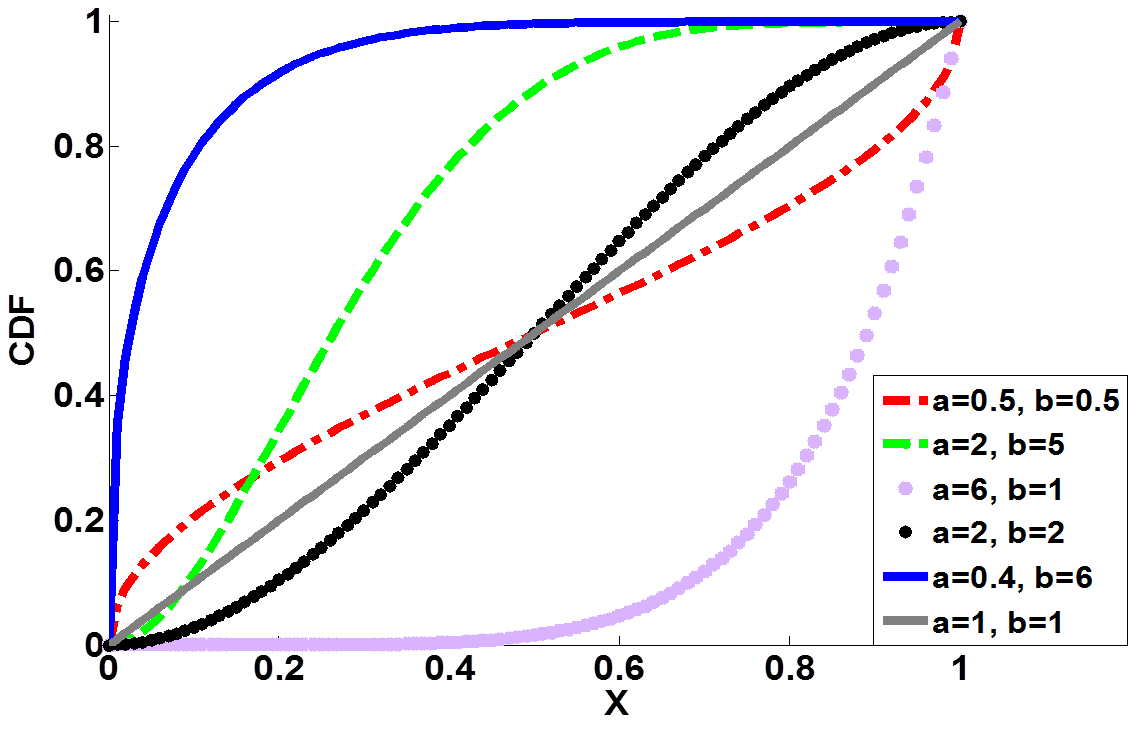}
  \caption{\label{betagraph} Beta-cdf plots for different values of the parameters $a$ and $b$}
\end{figure}

\begin{figure}[!ht]
\centering
  \includegraphics[width=0.75\textwidth]{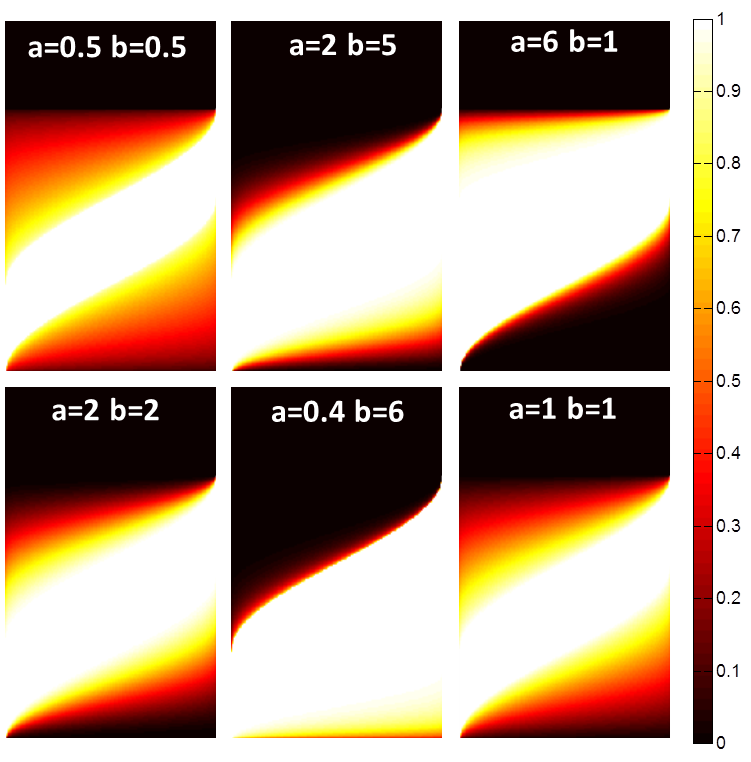}
  \caption{\label{betawt} Images showing weight distributions in fourier domain generated by parametric variation of the beta-cdf (using the parameters shown in Fig. \ref{betagraph}). Angular coverage is along abscissa and wavenumber varies from  $[-\nu_0, \nu_0]$}
\end{figure}

To this end, we use standard beta-cdf: $F(x|a,b) = \int_{-\infty}^x f(t|a,b)dt$, where $a>0$, $b>0$, and $f$ is the standard beta probability density function (see \cite{johnson1994continuous} \&   \cite{johnson1995continuous}):
\begin{equation}
f(t|a,b) =
\begin{cases}
  \frac{1}{B(a,b)} t^{a-1}(1-t)^{b-1}, &  \quad 0 \leq t \leq 1, \\
  0, & \quad \text{otherwise}.
\end{cases} \label{beta}
\end{equation}
and $B(a,b) = \int_0^1 t^{a-1}(1-t)^{b-1}dt$. We obtain family of beta-cdf's by changing values of $a$ and $b$ in (\ref{beta}), as shown in the Fig. \ref{betagraph}. Notice that, the standard beta-cdf has support $[0,1]$, while in region $A$ for a fixed $\nu_m$, we need to define weights for $\phi \in [0, 2\alpha+\pi/2]$. This can be achieved by substituting $x=\frac{\phi}{2\alpha+\pi/2}$ in the formula of $F$. The corresponding weight profiles in the frequency domain are shown in Fig. \ref{betawt}.

\begin{figure}
%\parbox[left]{
\begin{minipage}[t]{0.48\textwidth}
  \centering
%\raggedleft
  \includegraphics[width=0.95\linewidth, height=1.25\linewidth]{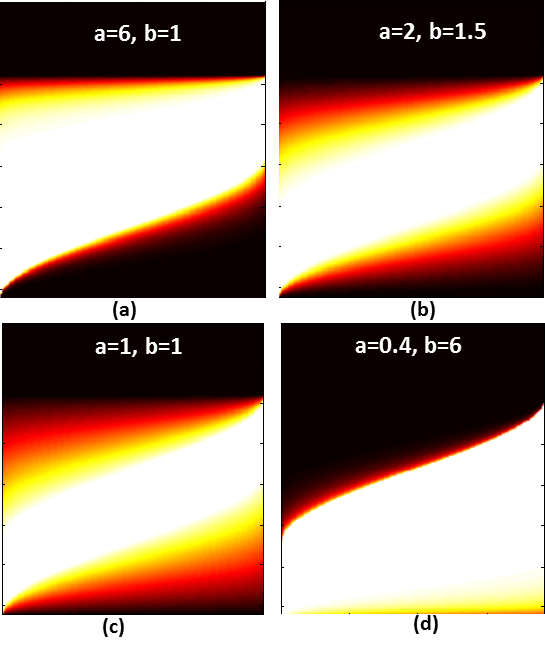}
  \captionof{figure}{Images showing weight distributions in Fourier domain generated by parametric variation of the beta-cdf. The parameters $a$ and $b$  used for each weight  distribution are inset in each plot. Frequency is plotted along abscissa and angular coverage along ordinate.}
  \label{betawt2}
\end{minipage} 
\quad %\quad % \qquad %\qquad
%\parbox[left]{
\begin{minipage}[t]{0.48\textwidth}
  \centering
%\raggedright
  \includegraphics[width=1\linewidth,height=1.25\linewidth]{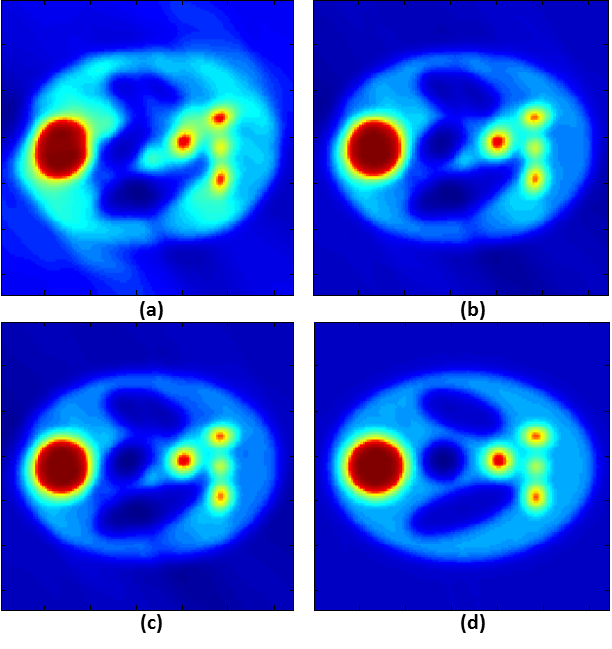}
  \captionof{figure}{Corresponding reconstructions from using the weight distributions in Fig. \ref{betawt2}}
  \label{betawt2_recon_eg}
\end{minipage}
\end{figure}
The plots in Fig. \ref{betagraph} can be used to understand how the weights will be distributed in Regions $A$ and $C$. Fig. \ref{betawt} gives useful insight towards choosing good parameters. For example,with combinations $a=2, b=5$ or $a=6, b=1$, region $A$ is not well covered, whereas for $a=0.4, b=6$, region $A$ has been almost fully covered with near unity weights leaving Region $C$ with mostly near-zero weights. This combination is expected to better utilize data redundancy than the other combinations. Especially for coverages lower than $270^{\circ}$, we receive less information from $C$, hence an optimum weight function should span most of region $A$ with near unit weightage. This can be further illustrated through Fig. \ref{betawt2} and Fig. \ref{betawt2_recon_eg}. Fig. \ref{betawt2} shows some weight distributions in the Fourier space (in the alternative co-ordinate system). The beta-cdf parameters used to generate these weights are shown in inset white text. Fig \ref{betawt2_recon_eg} shows the reconstructed real parts of the images by using these weights. Reconstruction was performed from sub-minimal angular coverage of $200^o$. The weights which cover region A more completely also give better reconstructions. Since the weights are continuous across the boundaries of the regions, the transition from zero to unity should be prompt enough to retain as much information as possible within region $A$ but also not too abrupt to generate artifacts in case of discrete datasets. To estimate optimum parameters, $a$ and $b$ were parametrically sweeped and the corresponding weight distributions in the frequency space were observed. The combination $a=0.4, b=6$ was found to be most effective overall, with the above mentioned constraints in view.

Similar approach can be used to generate weight functions from other cumulative distribution functions such as gamma-cdf, normal-cdf etc. However as both gamma and normal distributions have unbounded support, we need to either truncate or transform the arguments within the boundaries of region $A$. These distributions used along with their respective transformations are described in the Appendix. In this paper, results from different distributions using optimum parameters for each distribution will be presented and compared with the results obtained using weights given in (\ref{Panwt}).

\section{Results} \label{result}

\begin{figure}[!h]
\centering
  \includegraphics[width=0.65\textwidth]{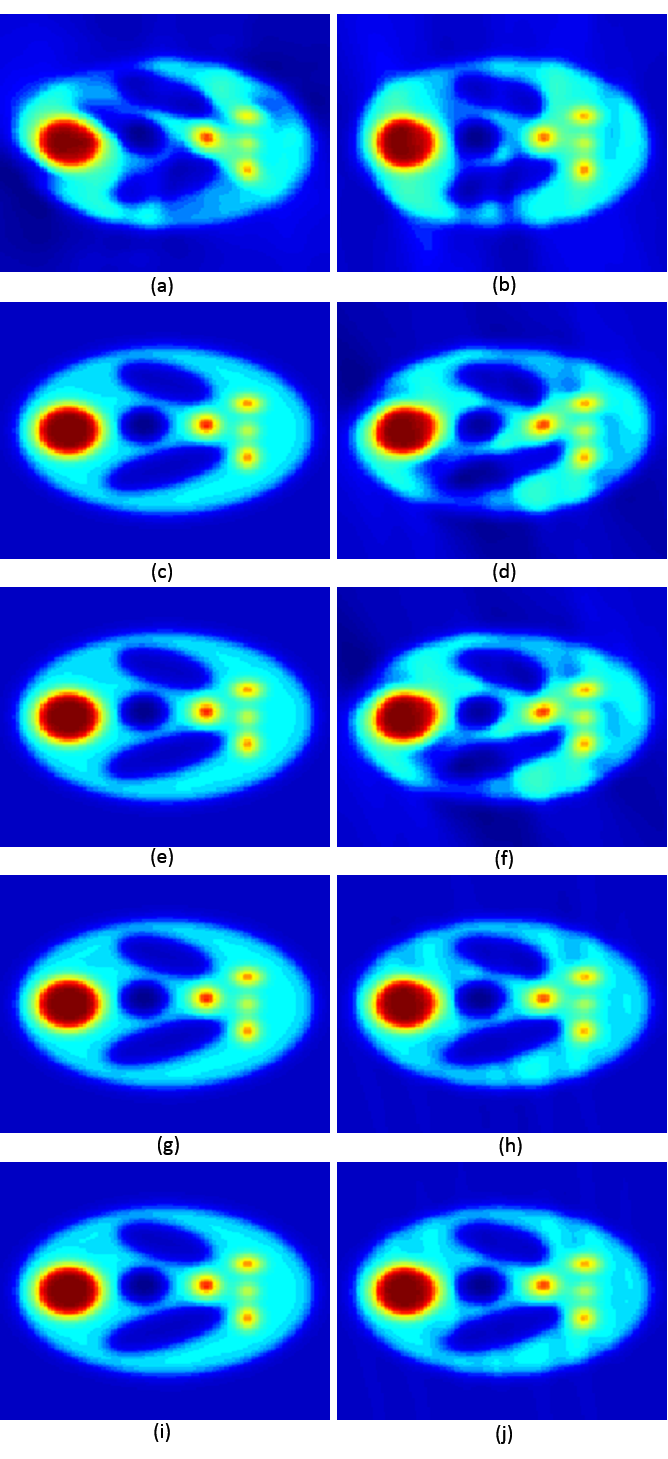}
  \caption{\label{bhoot1} Reconstructed real part from complex test image under noiseless conditions. The left column shows reconstruction from $270^{\circ}$ coverage and right column shows reconstruction from $200^{\circ}$ coverage. (a)-(b) using regular FBPP, (c)-(d) using sine-sq weights, (e)-(f) using Normal-cdf weights, (g)-(h) using gamma-cdf weights, (i)-(j) using beta-cdf weights}
\end{figure}

\begin{figure}[!h]
\centering
  \includegraphics[width=0.65\textwidth]{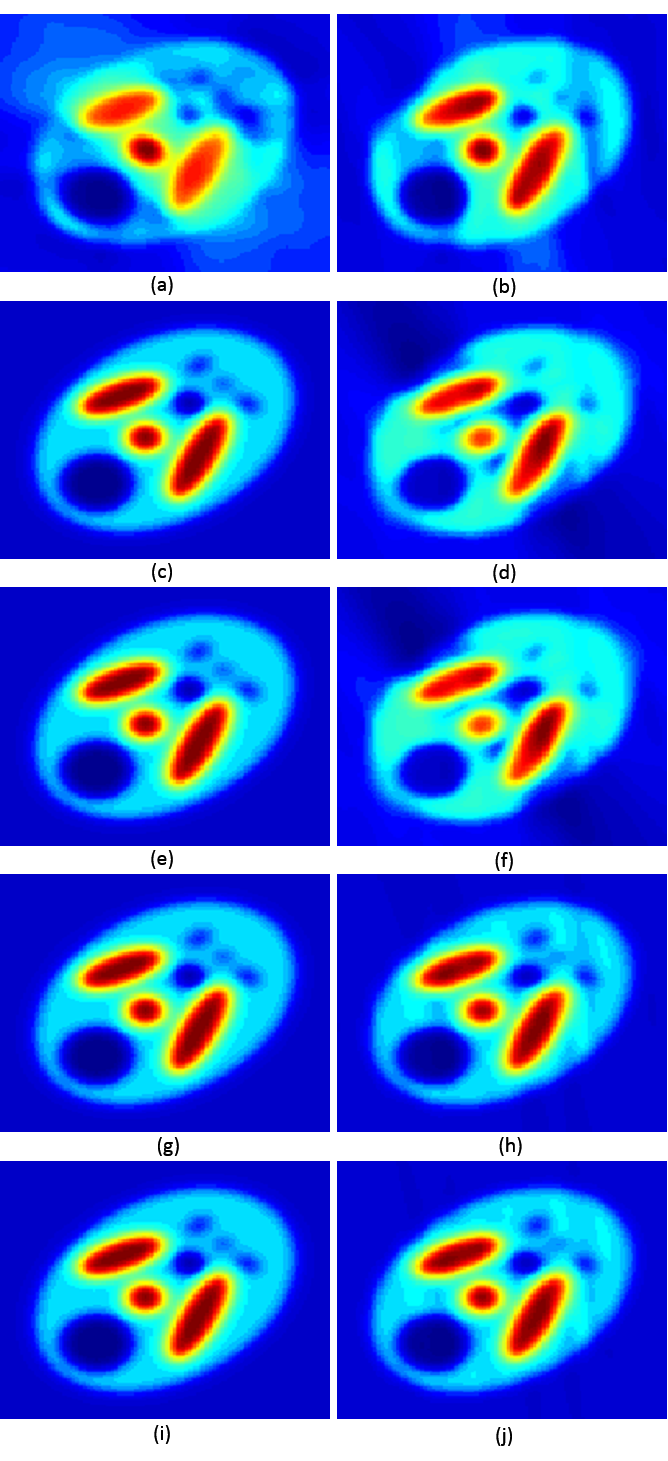}
  \caption{\label{bhoot2} Reconstructed imaginary part from complex test image under noiseless conditions. The left column shows reconstruction from $270^{\circ}$ coverage and right column shows reconstruction from $200^{\circ}$ coverage. (a)-(b) using regular FBPP, (c)-(d) using sine-sq weights, (e)-(f) using Normal-cdf weights, (g)-(h) using gamma-cdf weights, (i)-(j) using beta-cdf weights}
\end{figure}

This paper presents results from simulated projection data for DT valid under the Born approximation. Projection data has been calculated using analytical expressions from \cite{avinash2001principles}, \cite{pan1983computational} for the case of Shepp-Logan type phantoms. These equations can be used to calculate the Fourier transforms of diffracted projections from ellipses of specified foci and major and minor axes. The test image is a combination of multiple ellipses of different eccentricities located at different locations within the ROI and their major axes oriented at different angles. The image is complex. Both the real and imaginary part of the image are in effect, modified versions of the standard Shepp-Logan phantom, a standard model used to validate computed tomography algorithms. The real and imaginary parts of the test image are shown in Fig. \ref{fbppReal}(a) and \ref{fbppImg}(a). The image matrix is $128 \times 128$ pixels with pixel size of  $\lambda/8$. The image has an area of $16\lambda \times 16\lambda$. The objective of this paper was to find optimum weights which can exploit the redundancy for angular coverages below $270^{\circ}$ where redundancy is still present in the projection data. Proceeding in the manner described in the previous section to generate weights, the following optimum parameters for different cdfs were found: for beta-cdf, $a=0.4, b=6$; for gamma-cdf, $a=2.1, b=0.1$. For normal-cdf based weights, we used truncation method as explained in Appendix. For each $\nu_m$, we got the best result for $\mu=\tfrac{1}{2}(2\alpha+\pi/2)$, and $\sigma=\tfrac{1}{6}(2\alpha+\pi/2)$. Given below are reconstructed images from regular FBPP and MS-FBPP using different weight functions for coverages $270^{\circ}$ and $220^{\circ}$ (an example of sub-minimal angular coverage). Both real and imaginary parts are shown in Fig. \ref{bhoot1}  and \ref{bhoot2} respectively.

The reconstructions show the beta-cdf and gamma-cdf based weights generate a very stable reconstruction even in lower angular coverage of $200^{\circ}$, with beta-cdf performing slightly better overall. The sine-squared and Normal-cdf based reconstructions perform accurate reconstructions at $270^{\circ}$, but below that they progressively deteriorate and are clearly not optimum choices as seen at $200^{\circ}$. Below $180^{\circ}$ the redundancy disappears and using weights in principle cannot produce a better reconstruction than regular FBPP. So, reconstructed images are not shown for further lower coverage. However, for un-optimized weights, the reconstruction can actually worsen at a lower coverage. Ideally, the optimized filters would generate reconstructions similar to traditional FBPP at angles below $180^{\circ}$ but the un-optimized filters like the sine-squared filter would give a worse reconstruction than regular FBPP for these coverages.

\begin{figure}[!ht]
\centering
  \includegraphics[width=0.75\textwidth]{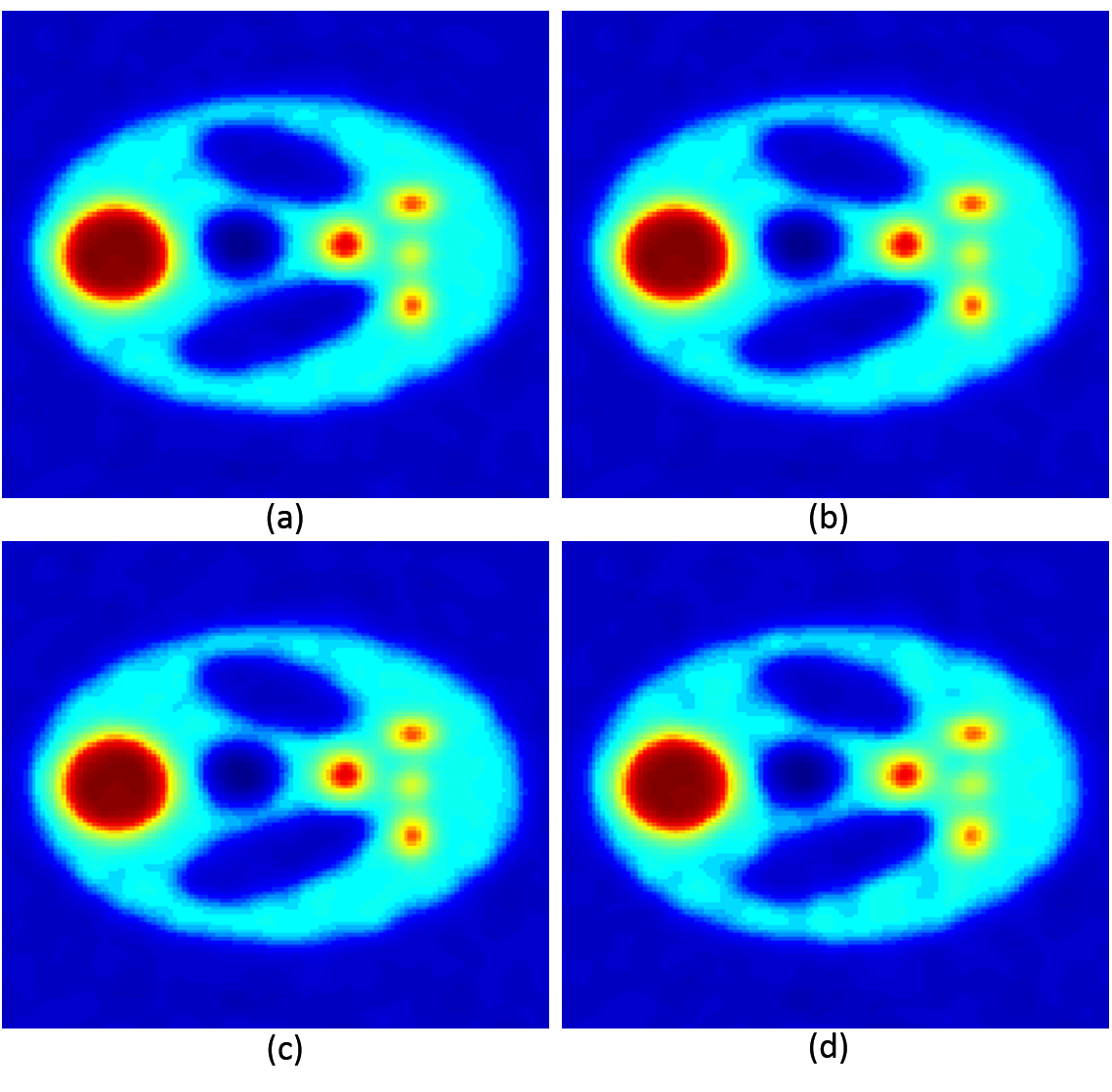}
  \caption{\label{bhoot3} Reconstruction of real part of image from 3dB awgn projection data using beta-cdf weights for angular coverages (a) $270^{\circ}$ (b) $250^{\circ}$ (c) $220^{\circ}$ (d) $200^{\circ}$}
\end{figure}

\begin{figure}[!ht]
\centering
  \includegraphics[width=0.75\textwidth]{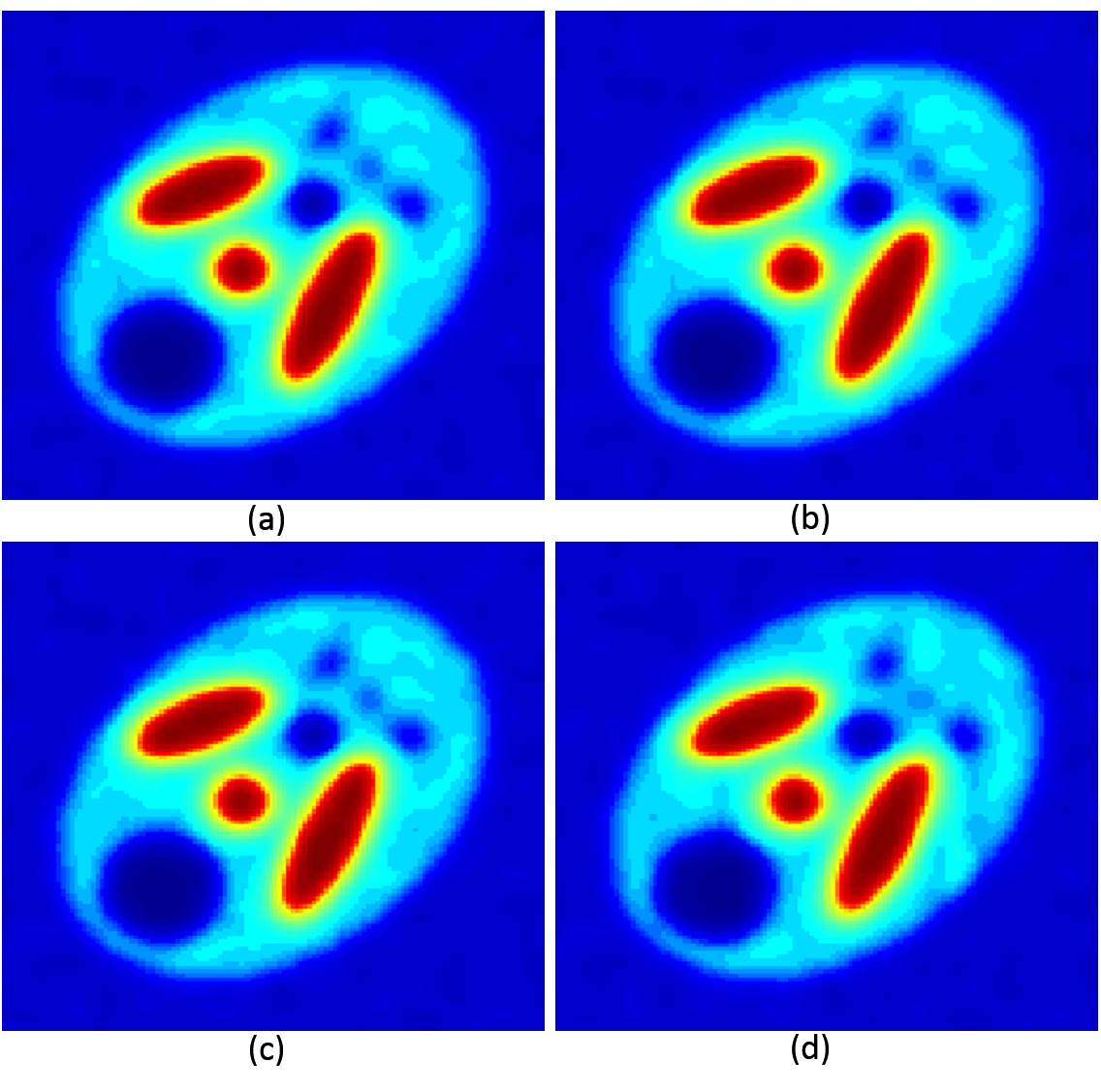}
  \caption{\label{bhoot4} Reconstruction of imaginary part of image from 3dB awgn projection data using beta-cdf weights for angular coverages (a) $270^{\circ}$ (b) $250^{\circ}$ (c) $220^{\circ}$ (d) $200^{\circ}$}
\end{figure}

Noise is an integral part of any measurement system. In a DT setup, noise may arise from random inhomogeneities in medium or as measurement noise introduced in the experimental procedure. To account for these, noisy reconstruction has been modeled as a stochastic process in literature before \cite{tsihrintzis1993stochastic}, \cite{anastasio2000computationally}, \cite{rouseff1991diffraction}. Noise analysis by itself is an involved field beyond the scope of this paper. The noisy data analysis was necessary to examine the reliability of these algorithms when applied to practical systems. The MS-FBPP algorithms are expected to respond to noisy data models differently. To consider the effect of all noise sources, it was assumed as sufficient to consider a white Gaussian noise distribution in the scattered field data \cite{pan1998unified}. An additive white Gaussian noise has been added to the analytically computed projection data for different variances to give different noise levels. To observe the non-uniform propagation of errors under noisy data \cite{pan1998unified}, \cite{anastasio1999investigation}, we compared the reconstruction from the weighted MS-FBPP algorithms with regular FBPP using a noisy data model. We present here, for demonstrative purposes, reconstructions from projection data injected with 3-dB additive white Gaussian noise (awgn). The reconstructed images from noisy data using beta-cdf weights for different angular coverages in the range of $[200^{\circ}, \thickspace 270^{\circ}]$ are given in Fig. \ref{bhoot3} and Fig. \ref{bhoot4}. We use here weights based on beta-cdf with $a=0.4$, $b=6$ in noisy condition, which earlier gave best result under noiseless condition. The reconstructions show robustness of the algorithm to noise levels that could be reasonably expected from good experimental data.

These show that the beta-cdf based weights are capable of maintaining all the features and without any artifacts up to $220^{\circ}$. The reconstruction at $200^{\circ}$ is also almost distortionless. The responses are stable and the images are not affected noticeably due to the noise injection.

\begin{figure}[!ht]
\centering
  \includegraphics[width=0.75\textwidth]{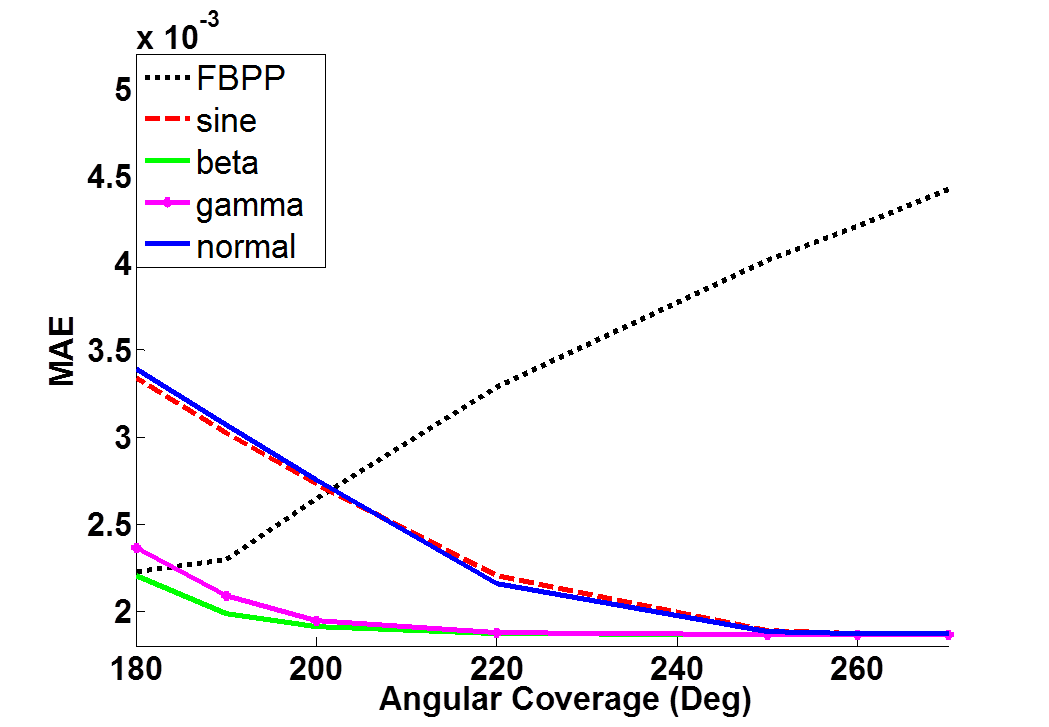}
  \caption{\label{MAE1}  MAE calculated at different coverages for real part of reconstructed image with regular FBPP and MS-FBPP using different weights}
\end{figure}

\begin{figure}[!ht]
\centering
  \includegraphics[width=0.75\textwidth]{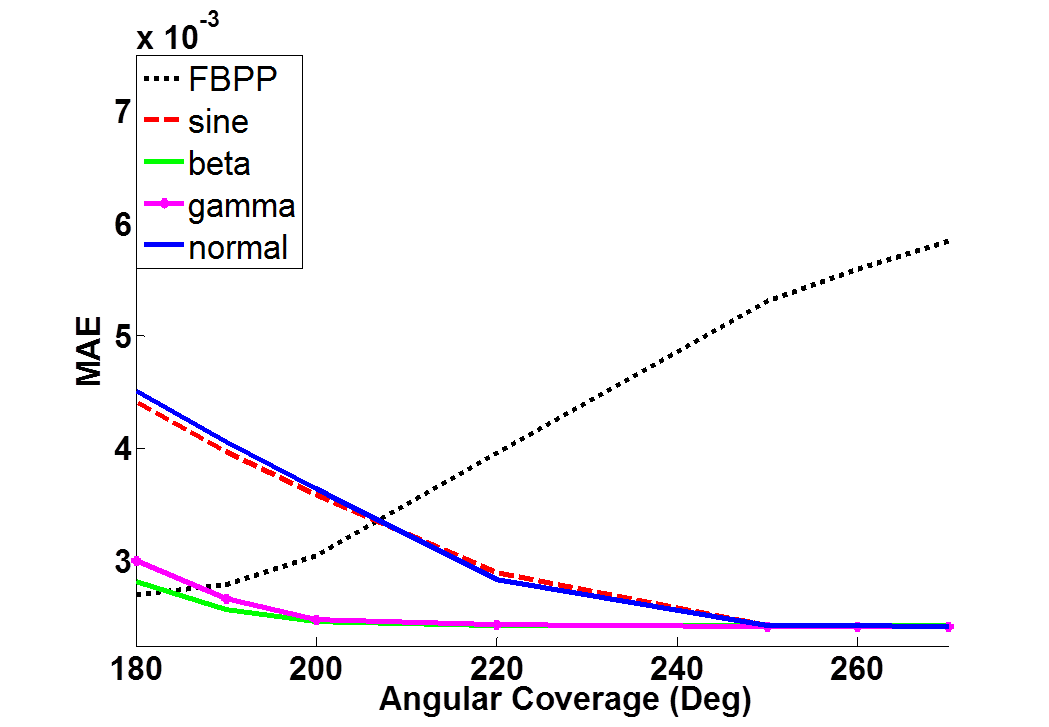}
  \caption{\label{MAE2} MAE calculated at different coverages for imaginary part of reconstructed image with regular FBPP and MS-FBPP using different weights}
\end{figure}

\begin{figure}[!ht]
\centering
  \includegraphics[width=0.75\textwidth]{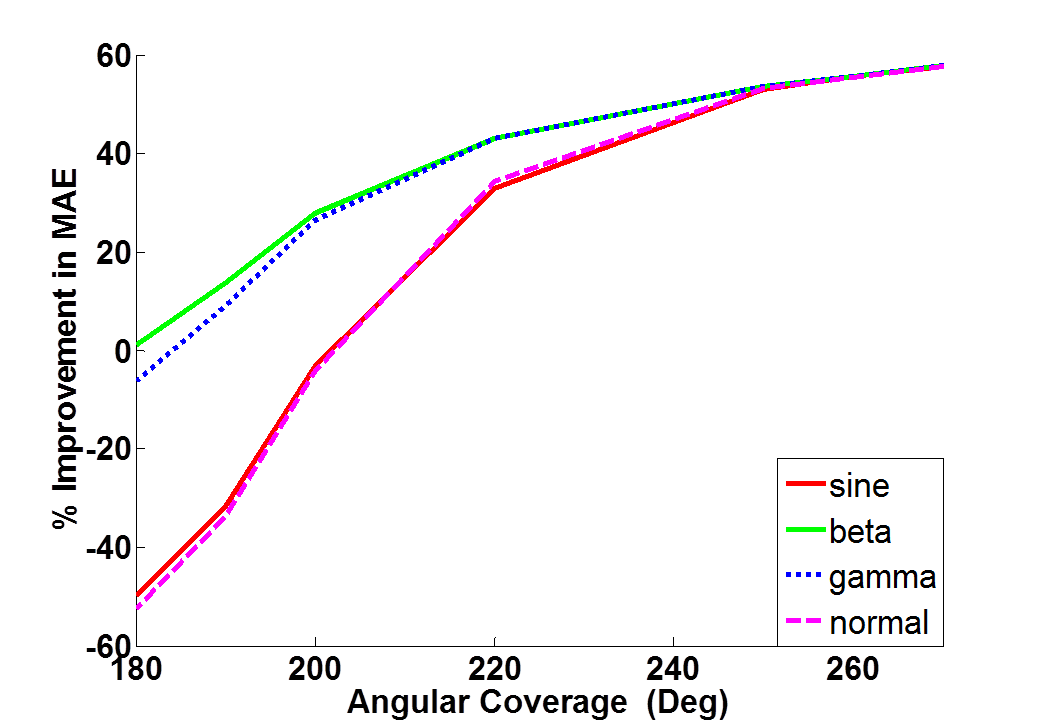}
  \caption{\label{MAEpct1} Percentage improvement of MAE from different weights over regular FBPP at different coverages for real part of reconstructed image}
\end{figure}

\begin{figure}[!ht]
\centering
  \includegraphics[width=0.75\textwidth]{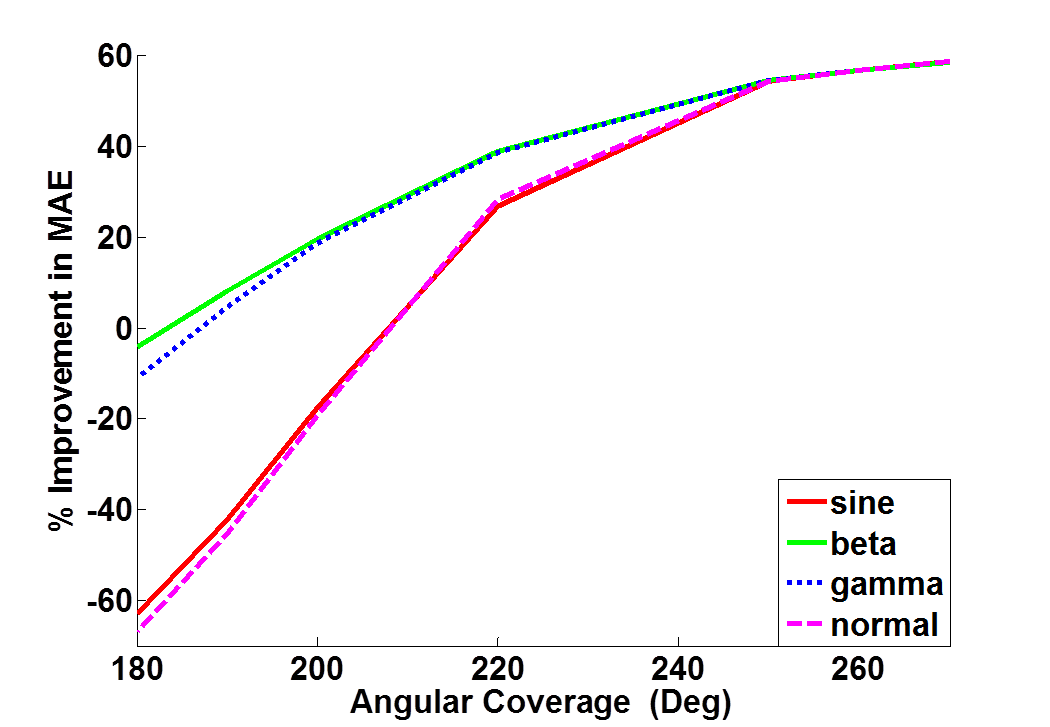}
  \caption{\label{MAEpct2} Percentage improvement of MAE from different weights over regular FBPP at different coverages for imaginary part of reconstructed image}
\end{figure}

To compare the performance of the different MS-FBPP algorithms we compare their Mean-Absolute-Error (MAE) with respect to the original image. The MAE is calculated as the absolute mean pixel-by-pixel difference between the original and reconstructed images. The real and imaginary parts of image were calculated separately.  The MAE calculated for the different classes at different angular coverages are plotted in Fig. \ref{MAE1} and Fig. \ref{MAE2}. The results show that as the coverage decreases below $270^{\circ}$, the beta-cdf and gamma-cdf weights are able to generate a constant MAE up to $200^{\circ}$. Beyond that, as the redundancy is lost, the MAE becomes comparable to a regular FBPP reconstruction. For the sine-squared and normal-cdf weights, the MAE is equal to the other two weights near $270^{\circ}$, but quickly rises as the coverage lowers. The error increases and is steadily greater than both the regular FBPP and other two MS-FBPP reconstructions. The percentage improvement of all the four classes with respect to the regular FBPP are plotted in Fig. \ref{MAEpct1} and \ref{MAEpct2}. All the filters show maximum improvement at $270^{\circ}$ and decreases with lower coverage. The degradation of non-optimal filters is much greater than optimal filters. For coverages below $180^{\circ}$, the optimal weights can still maintain a reconstruction equivalent to regular FBPP, whereas the non-optimal weights degrade more rapidly as they do not use the available projection data efficiently. These results show that properly chosen distribution functions can be used efficiently to exploit data redundancy for better reconstructions with sub-minimal angular coverages. Below $180^{\circ}$ coverage, an image reconstructed from optimum filter would also not degrade beyond a regular FBPP algorithm when the redundancy is lost due to highly limited angular coverage.

\section{Conclusion}

In this paper a generalized approach to use data redundancy within the traditional 2D DT setup has been explored. Using cumulative distribution functions especially the beta-cdf, it was shown that distortion-less reconstructions are possible even at angular coverage of $200^{\circ}$. The advantage of this observation is manifold. Firstly it reduces the angular requirements for accurate reconstructions. This also implies shorter access times for collecting relevant projection data. In medical applications this can also mean lower amount of exposure to the interrogating radiation. Results have been validated through simulated data. The performance of these algorithms with real experimental data will be explored in the future work of this study. For still lower coverage, the redundancy in the tomographic dataset vanishes and these algorithms cannot maintain accurate reconstructions below $180^{\circ}$. For such coverage, various iterative optimization techniques like total variation (TV) minimization, compressed sensing etc are being explored for application to DT setups. Combining these techniques with MS-FBPP could be a scope of future research work.

\appendix

%\section{Weight generation through gamma and normal cdf}

\noindent\textbf{Gamma-cdf based weighting:} for a fixed $\nu_m$, we define $y_{\phi} = \tan\left(\frac{\pi}{2}\frac{\phi}{2\alpha+\pi/2}\right)$. Then
\begin{equation}
w(\nu_m, \phi) = \frac{1}{\Gamma(a)b^a} \int_{-\infty}^{y_{\phi}} t^{a-1}e^{-t/b} dt.
\end{equation}
where $\Gamma(a) = \int_0^{\infty} t^{a-1}e^{-t}dt$, and $a>0$, $b>0$ are parameters. \medskip

\noindent\textbf{Normal-cdf based weighting:} In case of normal-cdf, we used truncation method as follows:
\begin{equation}
  w(\nu_m, \phi) = \frac{F(\nu_m|\mu,\sigma)-F(0|\mu,\sigma)}{F(2\alpha+\pi/2|\mu,\sigma)-F(0|\mu,\sigma)},
\end{equation}
where $\mu\in (-\infty, \infty)$ and $\sigma >0$ are parameters and
\begin{equation}
F(x|\mu,\sigma) = \frac{1}{\sqrt{2\pi}\sigma} \int_{-\infty}^x e^{-(t-\mu)^2/2\sigma^2}dt.
\end{equation}

%%%%%%%%%%%%%%%%%%%%%%%%%%%%%%%%%%%%

\end{document}